\documentstyle[12pt,a4]{article}
\newcommand{\beq}{\begin{equation}}
\newcommand{\eeq}{\end{equation}}
\newcommand{\beqa}{\begin{eqnarray}}
\newcommand{\eeqa}{\end{eqnarray}}
\newcommand{\beqan}{\begin{eqnarray*}}
\newcommand{\eeqan}{\end{eqnarray*}}
\newcommand{\no}{\nonumber}

\newcommand{\lets}{\stackrel{<}{\sim}}

\newcommand{\ol}{\overline}
\newcommand{\ra}{\rightarrow}

\newcommand{\ve}{\varepsilon}

\newcommand{\dg}{\dagger}

\newcommand{\wh}{\widehat}
\newcommand{\dfrac}{\displaystyle \frac}
\newcommand{\dis}{\displaystyle}
\newcommand{\ben}{\begin{enumerate}}
\newcommand{\een}{\end{enumerate}}
\newcommand{\bfl}{\begin{flushleft}}
\newcommand{\efl}{\end{flushleft}}
\newcommand{\ba}{\begin{array}}
\newcommand{\ea}{\end{array}}
\newcommand{\btab}{\begin{tabular}}
\newcommand{\etab}{\end{tabular}}
\newcommand{\bit}{\begin{itemize}}
\newcommand{\eit}{\end{itemize}}

\newcommand{\Fsl}{\not\!\!}

\newcommand{\cL}{{\cal L}}

\normalsize
\begin{document}
\parskip=4pt plus 1pt  
\begin{titlepage}
\begin{flushright}
UWThPh-1993-31\\
September 1993
\end{flushright}
\vspace{1.5cm}
\begin{center}
{\Large \bf  THE STANDARD MODEL AT LOW ENERGIES*}\\[40pt]
{\bf G. Ecker} \\[5pt]
Institut f\"ur Theoretische Physik \\
Universit\"at Wien \\
Boltzmanngasse 5, A-1090 Wien
\vfill
{\bf Abstract} \\[10pt]
\end{center}
\noindent
The hadronic sector of the standard model at low energies is described
by a non--decoupling effective field theory, chiral perturbation
theory. An introduction is given to the construction of
effective chiral Lagrangians, both in the purely mesonic sector and
with inclusion of baryons. The connection between the relativistic
formulation and the heavy baryon approach to chiral perturbation
theory with baryons is reviewed.

\vfill
\begin{center}
Lectures given at the \\[3pt]
$6^{\rm th}$ Indian--Summer School on Intermediate Energy Physics \\[3pt]
Interaction in Hadronic Systems \\[3pt]
Prague, August 25 - 31, 1993 \\[3pt]
To appear in the Proceedings (Czech. J. Phys.)
\end{center}

\vfill
\noindent * Work supported in part by FWF, Project No. P09505-PHY
(EURODA$\Phi$NE Collaboration)
\end{titlepage}
\tableofcontents

\renewcommand{\theequation}{\arabic{section}.\arabic{equation}}
\setcounter{equation}{0}
\section{EFFECTIVE FIELD THEORIES}
Until some 25 years ago, quantum field theory as a tool for making
experimentally testable predictions in particle physics seemed
to be restricted to quantum electrodynamics. The $S$--matrix
approach was widely considered to be a more promising method
for the strong and even the weak interactions. The development and
the impressive success of the standard model of strong and electroweak
interactions have led to a grand revival of quantum field theories.
In the more recent past, so-called effective field theories
(EFT) have
been used more and more frequently. In a certain sense, EFT can be
regarded as a synthesis of $S$--matrix theory and quantum field theory
\cite{Wein92}.
Phrased in a different manner, Lagrangian quantum field theory offers
an explicit framework for implementing systematically all
the general axioms of $S$--matrix
theory like analyticity, unitarity and cluster decomposition for
a finite number of degrees of freedom \cite{Wein79} (the revenge of
$S$--matrix theory according to Weinberg \cite{Wein92}). Even our present
``fundamental'' quantum field theories are in all likeliness just
EFT below some energy scale. All we know
about this scale is that it must be  larger than the presently
available energies.

The notion of EFT is only practicable if we have, in addition to the general
axioms of $S$--matrix theory,
some other criteria at our disposal  to constrain the corresponding effective
Lagrangians. These criteria are the symmetries we abstract from the
``fundamental'' underlying theory, Lorentz invariance being the most
obvious one. In this context, it is useful to distinguish between
two types of EFT.

\subsection*{\bf A. Decoupling EFT}
For energies below some scale $\Lambda$, all heavy (with respect
to $\Lambda$) degrees of freedom are integrated out leaving only
the light degrees of freedom in the effective theory. No light
particles are generated in the transition from the fundamental
to the effective level. The effective Lagrangian has the general
form
\beq
\cL_{\rm eff} = \cL_{d\leq 4} + \sum_{d>4} \frac{1}{\Lambda^{d-4}}
\sum_{i_d} g_{i_d} O_{i_d}  \label{eft}
\eeq
where $\cL_{d\leq 4}$ contains the potentially renormalizable terms
with operator dimension $d \leq 4$, the $g_{i_d}$ are dimensionless
coupling constants, expected to be at most of $O(1)$,
and the $O_{i_d}$ are monomials
in the light fields with operator dimension $d$. At energies
much below $\Lambda$, it may be a good approximation to work
with the renormalizable Lagrangian $\cL_{d\leq 4}$ only, because the
corrections due to the non--renormalizable parts ($d>4$) will be suppressed
by powers of $E/\Lambda$. In such cases, $\cL_{d\leq 4}$ can be regarded
as the ``fundamental'' Lagrangian at low energies.

\noindent
There are several well--known examples of decoupling EFT.
\ben
\item[a.] {QED for $E \ll m_e$} \\
For photon energies much smaller than the electron mass, the electrons
can be integrated out to yield the Euler--Heisenberg Lagrangian
for light-by-light scattering \cite{EuHe}
\beq
\cL_{\rm EH} = \frac{1}{2}(\vec{E}^2 - \vec{B}^2) + \dfrac{e^4}
{360\pi^2 m_e^4} \left [(\vec{E}^2 - \vec{B}^2)^2 + 7 (\vec{E}\cdot
\vec{B})^2 \right] + \dots \label{EH}
\eeq
The coefficient in front of the $d=8$ interaction term can be
calculated explicitly from the box diagram, because the ``fundamental''
theory QED can be treated perturbatively. Here, the renormalizable
part is only the kinetic term.
\item[b.] {Weak interactions for $E \ll M_W$} \\
At low energies, the weak interactions reduce
to the Fermi theory including the neutral current
interactions. After integrating out the $W$ and $Z$ bosons, the weak
interactions appear first in the $d=6$ terms.
\item[c.] {The standard model for $E \ll 1 \mbox{ TeV}$} \\
There is no reason to believe that the standard model
will be valid down to arbitrarily small distances. Instead, it should
be regarded as the EFT of some more fundamental
theory at higher energies (SUSY, grand unification, superstrings,\dots).
At presently available energies, however, all experimental evidence
is still in agreement with the standard model Lagrangian
being ``fundamental''.
\een
\subsection*{\bf B. Non--decoupling EFT}
In the transition from the fundamental to the effective level, a
phase transition occurs via the
spontaneous breakdown of some symmetry generating light
($M \ll \Lambda$) pseudo--Goldstone bosons. Since a spontaneously broken
symmetry relates processes with different numbers of Goldstone bosons,
a split of the effective Lagrangian like in Eq. (\ref{eft}) into
renormalizable ($d\leq 4$) and non--renormalizable ($d>4$) parts
becomes meaningless. The effective Lagrangian of a non--decoupling EFT is
intrinsically non--renormalizable. As the developments of the last
decade have shown \cite{Wein79,GL1,GL2},
this does not prevent such EFT from being perfectly
consistent quantum field theories. Instead of the operator dimension
like in the effective Lagrangian (\ref{eft}), it
is the number of derivatives of the fields that distinguishes
successive terms in the Lagrangian.

The general structure of effective Lagrangians with spontaneously broken
symmetries is to a large extent independent of the specific physical
realization. This can already be seen from two examples
in particle physics.
\ben
\item[a.] {The standard model without Higgs bosons} \\
Even if there is no explicit Higgs boson or if its mass is
bigger than about 1 TeV, the gauge symmetry $SU(2)\times U(1)$
can be spontaneously broken to $U(1)_{\rm em}$ (see, e.g., Ref. \cite{HH}
for a recent review of the heavy Higgs scenario). Since it is a local
symmetry that is spontaneously broken, the Goldstone bosons are not
physical particles, but they become the longitudinal components of the
$W$ and $Z$ bosons. As a manifestation of the universality of Goldstone
boson interactions, the scattering of longitudinal vector bosons
is in first approximation completely analogous to $\pi\pi$ scattering.
\item[b.] {The standard model for $E< 1 \mbox{ GeV}$} \\
At low energies, the relevant degrees of freedom of the standard
model are not quarks and gluons, but the pseudoscalar mesons
(and eventually other hadrons),
the pseudo-Goldstone bosons of spontaneously broken chiral
symmetry. The standard model in the hadronic sector is
described by a non--decoupling EFT  called chiral
perturbation theory (CHPT).
\een

\noindent
The applications of EFT can again be classified in two groups:
\bit
\item If the fundamental theory is unknown, EFT can be used to parametrize
the effects of new physics. Examples are the heavy Higgs
scenario and the search for effects of grand unification.
\item Even if the fundamental theory is known, EFT can be useful.
Either the full structure of the underlying theory is not needed
as for light-by-light scattering, or the fundamental theory
is not applicable with reliable methods in the
region of interest. The latter case applies to QCD in the confinement
regime where CHPT takes over.
\eit

In Sect. 2, a short introduction to the basic ingredients of
CHPT will be given. The effective chiral Lagrangian for the
pseudoscalar mesons up to $O(p^4)$ will be discussed in Sect. 3.
Some recent applications of CHPT in the mesonic sector are
briefly summarized. Finally, in Sect. 4 both the relativistic and the
so--called non--relativistic formulation of baryon CHPT are reviewed.
As an interesting application, neutral pion photoproduction off nucleons
at threshold is treated in some detail.

In addition to the references given during the course of the lectures,
more detailed material both on the structure and on
applications of CHPT can be found in Refs.~[7 -- 15].

\renewcommand{\theequation}{\arabic{section}.\arabic{equation}}
\setcounter{equation}{0}
\setcounter{section}{1}
\section{CHIRAL PERTURBATION THEORY}
CHPT is the EFT of the standard model at low energies in the
hadronic sector. Since as an EFT it contains {\bf all} terms
allowed by the symmetries of the underlying theory \cite{Wein79},
it should not be viewed as a particular model for hadrons, but rather
as a direct consequence of the standard model itself. The main
assumptions of CHPT are:
\bit
\item The masses of the light quarks $u,d$ and possibly $s$ can
be treated as perturbations.
\item In the limit of zero quark masses,
the resulting chiral symmetry is spontaneously
broken to its vectorial subgroup, isospin ($n=2$) or flavour $SU(3)$
($n=3$) for $n$ massless quarks. The resulting Goldstone
bosons are the pseudoscalar mesons.
\eit

CHPT is a systematic expansion around the chiral limit \cite
{Wein79,GL1,GL2}. As an EFT, it is a low--energy expansion in momenta
(and pseudoscalar meson masses). Here are some of its prominent features:
\bit
\item It is a well--defined
quantum field theory even though it is non--renormalizable.
\item There is no problem of double--counting of degrees of freedom:
the effective chiral Lagrangian contains only hadron fields,
but neither quarks nor gluons.
\item The softly broken chiral symmetry and electromagnetic gauge
invariance are manifest. The corresponding
Ward identities are automatically satisfied.
\item CHPT is ideally suited for the physics of pseudoscalar
mesons, but other hadrons can be coupled in a chiral invariant
manner (cf. Sect. 4 for the inclusion of baryons).
\item The chiral anomaly \cite{ABJB} is unambiguously incorporated
at the hadronic level \cite{WZW}.
\item Although the structure of Green functions and amplitudes is
calculable in CHPT, they contain parameters which are not restricted
by the symmetries of the standard model. At each order in the chiral
expansion, those low--energy constants must be determined either
from phenomenology or with additional model--dependent assumptions.
\eit

\subsection{Non--Linear Realization of Chiral Symmetry}
The QCD Lagrangian for $n$ massless quarks $q=(u,d,\dots)$
\beqa
{\cL}^0_{\rm QCD} &=& \ol{q} i \gamma^\mu\left(\partial_\mu + i g_s {\lambda_
\alpha\over 2} G^\alpha_\mu\right)q - {1\over 4}G^\alpha_{\mu\nu}
G^{\alpha\mu\nu} + \cL_{\mbox{\tiny heavy quarks}} \label{QCD}\\*
&=& \ol{q_L} i \Fsl{D} q_L + \ol{q_R} i \Fsl{D} q_R + \dots  \no \\*
q_{R,L} &=& {1\over 2}(1 \pm \gamma_5)q \no
\eeqa
has a global symmetry
$$
\underbrace{SU(n)_L \times SU(n)_R}_{\mbox{chiral group $G$}}
\times U(1)_V \times U(1)_A ~.
$$
All theoretical and phenomenological evidence suggests that the
chiral group $G$ is spontaneously broken to the vectorial subgroup
$SU(n)_V$. The axial generators of $G$ are non--linearly realized
and there are $n^2 - 1$ massless pseudoscalar Goldstone bosons.

There is a well--known procedure \cite{CCWZ} how to realize a
spontaneously broken symmetry on quantum fields. In the special case
of chiral symmetry with its parity transformation, the Goldstone
fields can be collected in a unitary matrix field $U(\phi)$ transforming as
\beq
U(\phi) \stackrel{g}{\ra} g_R U(\phi) g_L^{-1} \qquad
g = (g_L,g_R) \in G \label{Utr}
\eeq
under chiral rotations. There are different parametrizations of $U(\phi)$
corresponding to different choices of coordinates for the chiral
coset space $SU(n)_L\times SU(n)_R / SU(n)_V$. A
convenient choice is the exponential parametrization (for $n=3$)
\beq
U(\phi)=\exp{(i\lambda_a \phi^a/F)} \qquad
\Phi = {1 \over \sqrt{2}}\lambda_a \phi^a = \left( \ba{ccc}
\dfrac{\pi^0}{\sqrt{2}} + \dfrac{\eta_8}{\sqrt{6}} & \pi^+ &  K^+ \\*
\pi^- & -\dfrac{\pi^0}{\sqrt{2}} + \dfrac{\eta_8}{\sqrt{6}} &  K^0 \\*
K^- & \ol{K^0} & - \dfrac{2 \eta_8}{\sqrt{6}} \ea \right)~,
\eeq
where $F$ will turn out to be the pion decay constant in the
chiral limit. A more basic quantity from a geometrical point of view
is another matrix field $u(\phi)$ which in the standard choice
of coset coordinates is just the square root of $U(\phi)$
\cite{CCWZ}. Its chiral transformation
\beq
u(\phi) \stackrel{g}{\ra} g_R u(\phi) h(g,\phi)^{-1}
= h(g,\phi) u(\phi) g_L^{-1}
\eeq
introduces the so--called compensator field $h(g,\phi)$ representing an
element of the conserved subgroup $SU(n)_V$. For $g\in SU(n)_V$,
i.e. for $g_L = g_R$, the compensator $h(g)$ is a usual unitary representation
matrix, independent of the Goldstone fields $\phi$ (linear
representation \`a la Wigner--Weyl). For a proper chiral transformation
($g_L\not= g_R$), on the other hand, $h(g,\phi)$ does depend on $\phi$
(non--linear realization \`a la Nambu--Goldstone).

One may wonder why we should bother to introduce the matrix $u(\phi)$
with its concomitant compensator field, if we can construct the effective
chiral Lagrangian with the square $U=u^2$ with its simpler
transformation rule (\ref{Utr}). In fact, in the mesonic sector the
choice of representation is merely a matter of taste. Let us construct
as an exercise the lowest--order non--trivial chiral invariant
with Goldstone fields only. As a consequence of the Goldstone theorem,
the only such invariant without derivatives is
a trivial constant. Instead of the derivative $\partial_\mu U$ we can
also consider the so--called vielbein field
\beq
u_\mu = i (u^\dg \partial_\mu u - u \partial_\mu u^\dg) =
i u^\dg \partial_\mu U u^\dg ~,
\eeq
which transforms as
\beq
u_\mu \stackrel{g}{\ra} h u_\mu h^{-1}
\eeq
under chiral transformations. Since the chiral symmetry is
global for the time being, $\partial_\mu U$ transforms like $U$
itself in Eq. (\ref{Utr}). The unique chiral invariant of lowest
order, $O(p^2)$, can now be written in two equivalent ways as
($\langle \dots \rangle$ denotes the trace in $n$--dimensional flavour space)
\beq
\langle u_\mu u^\mu\rangle =
\langle\partial_\mu U^\dg \partial^\mu U\rangle~,\label{sig}
\eeq
which is nothing but the non--linear $\sigma$ model up to a constant
factor.

The representation involving the compensator field $h(g,\phi)$
is more convenient if we want to couple non--Goldstone degrees of
freedom like meson resonances (Sect. 3) or baryons (Sect. 4) (see
Ref. \cite{Georgi} for a discussion of different possibilities how to
represent the baryons in chiral Lagrangians). Once we specify the
transformation behaviour of (non--Goldstone) matter fields $\Psi$
under the flavour group $SU(n)_V$, it is straightforward to construct
a realization of the whole chiral group $G$ \cite{CCWZ}. Let us
take the nucleon doublet ($n = 2$) $\Psi$ as an example. Under
isospin it transforms as
\beq
\Psi \stackrel{g}{\ra} \Psi' = h(g) \Psi \quad \qquad g\in SU(2)_V
\eeq
where $h(g)$ is just the defining representation of $SU(2)$. For a
general chiral transformation $g \in G$ one defines
\beq
\Psi \stackrel{g}{\ra} \Psi' = h(g,\phi) \Psi~ \label{Ptr}
\eeq
in terms of the compensator field $h(g,\phi)$ in the doublet
representation \footnote{Of course, the same procedure works for
arbitrary representations of $SU(n)$.}. It is easy
to verify that (\ref{Ptr}) is a realization of $G$.
However, since the transformation (\ref{Ptr}) is local via
$h(g,\phi(x))$, the normal derivative $\partial_\mu\Psi$ does not
transform covariantly. The necessary connection is given by
\beq
\Gamma_\mu = \frac{1}{2} ( u^\dg \partial_\mu u
+ u \partial_\mu u^\dg)~.
\eeq
The corresponding covariant derivative
\beq
\nabla_\mu \Psi = \partial_\mu \Psi + \Gamma_\mu \Psi
\eeq
transforms indeed like $\Psi$ itself. In this framework, the
construction of chiral invariant Lagrangians is almost trivial.
Similar to gauge theories, it boils down to replacing normal by
covariant derivatives in $SU(n)_V$ invariant Lagrangians. Of course,
the recipe is so simple because the Goldstone fields appear only
in the matrix field $u(\phi)$.

\renewcommand{\theequation}{\arabic{section}.\arabic{equation}}
\setcounter{equation}{0}
\setcounter{section}{2}
\section{MESONS}

\subsection{Effective Chiral Lagrangian}
We first extend the QCD Lagrangian (\ref{QCD}) by
coupling the quarks to external hermitian matrix--valued fields
$v_\mu,a_\mu,s,p$:
\beq
{\cL} = {\cL}^0_{\rm QCD} + \bar q \gamma^\mu (v_\mu + \gamma_5 a_\mu)q
- \bar q (s - i \gamma_5 p)q ~. \label{SM}
\eeq
The external field method offers several advantages \cite{GL1,GL2}.
Its two most important features are:
\bit
\item The electromagnetic and semileptonic weak interactions are
automatically included with ($n=3$ for the rest of this section)
\beqa
r_\mu  =  v_\mu + a_\mu & = & - eQ  A_\mu
\label{gf} \\*
l_\mu  =  v_\mu - a_\mu & = &  - eQ  A_\mu
        - \dfrac{e}{\sqrt{2}\sin{\theta_W}} (W^+_\mu T_+ + h.c.) \no
\eeqa $$
Q = \dfrac{1}{3} \mbox{diag}(2,-1,-1)~, \qquad
T_+ = \left( \ba{ccc}
0 & V_{ud} & V_{us} \\
0 & 0 & 0 \\
0 & 0 & 0 \ea \right)  $$
where the $V_{ij}$ are Kobayashi--Maskawa matrix elements.
\item
The method provides an elegant way to incorporate
explicit chiral symmetry breaking through the quark masses. The
physical Green functions (and $S$--matrix elements) are obtained as
functional derivatives of a generating functional $Z[v,a,s,p]$
to be defined below at
$$
  v_\mu = a_\mu = p = 0~,
$$
but at
\beq
s = {\cal M} = \mbox{diag} (m_u,m_d,m_s)~.
\eeq
The tremendous advantage is that we can calculate
$Z[v,a,s,p]$ in a manifestly
chiral invariant way until the very end.
\eit
The previous global chiral symmetry is now promoted to a
local $SU(3)_L\times SU(3)_R$ symmetry
\beqa
q_{R,L} & \ra & g_{R,L} q_{R,L} \no \\*
r_\mu  & \ra & g_R r_\mu g^\dagger_R +
                             i g_R \partial_\mu g^\dagger_R \no \\*
l_\mu  & \ra & g_L l_\mu g^\dagger_L +
                             i g_L \partial_\mu g^\dagger_L \\*
s + i p & \rightarrow & g_R (s + i p) g^\dagger_L \no \\*
g_{R,L} & \in & SU(3)_{R,L}~. \no \eeqa
The local nature of the chiral symmetry requires the introduction of
a covariant (now with respect to the external gauge fields $v,a$)
derivative
\beq
D_\mu U = \partial_\mu U - ir_\mu U + iU l_\mu
\eeq
and of non--Abelian field strength tensors
\beqa
  F_R^{\mu\nu} & = & \partial^\mu r^\nu -
                     \partial^\nu r^\mu -
                     i [r^\mu,r^\nu]   \\*
  F_L^{\mu\nu} & = & \partial^\mu l^\nu -
                     \partial^\nu l^\mu -
                     i [l^\mu,l^\nu] ~. \no
\eeqa

In the mesonic sector, the generating functional $Z[v,a,s,p]$
can be defined via
\beq
e^{\dis i Z[v,a,s,p]} = N' <0|T e^{\dis i \int d^4x \cal L}|0> =
N \int [dU(\phi)] e^{\dis i \int d^4x \cL_{\rm eff}}~, \label{master}
\eeq
where the normalization constants $N, N'$ ensure $Z[0,0,0,0]=0$.
The first equation is the general definition of a generating functional
of Green functions in terms of the fundamental Lagrangian (\ref{SM}).
The content of the second
equation is that we shall perform the actual calculations
with the EFT of the standard model characterized by an effective chiral
Lagrangian
\beq
\cL_{\rm eff} = \cL_M = \cL_2 + \cL_4 + \dots~,\label{LMeff}
\eeq
written as an expansion in derivatives and external fields.
The chiral counting rules are ~:
\beqa
U & \hspace{3cm} & O(p^0) \no \\*
D_\mu U, v_\mu, a_\mu & & O(p) \no \\*
s,p,F_{L,R}^{\mu\nu} & & O(p^2) \label{cc}
\eeqa
which are obvious except for $s,p$, to be explained shortly.

The lowest order (locally) chiral invariant Lagrangian
\beq
{\cL}_2 = \frac{F^2}{4} \langle D_\mu U D^\mu U^\dagger +
             \chi U^\dagger + \chi^\dagger U \rangle \label{L2M} \eeq
is the generalization of the non--linear $\sigma$ model (\ref{sig})
in the presence of external fields with
\beq
\chi = 2 B_0(s + ip)~.\eeq
The parameters $F$ and $B_0$ are the only free
constants at $O(p^2)$. From a calculation of the axial--vector current
for the Lagrangian (\ref{L2M}), one finds
that $F$ equals the pion decay constant to lowest order in
CHPT~:
\beq
F=F_\pi=93.2 \mbox{ MeV}~.
\eeq
The second constant $B_0$ is related to the quark condensate~:
\beq
\langle 0|\ol{q_i}q_j|0\rangle =
- F^2 B_0 \delta_{ij}~. \label{qcond}
\eeq
It sets the overall scale for the pseudoscalar meson masses
at lowest order, e.g.
\beq
M^2_{\pi^+}  =  (m_u + m_d) B_0~.
\eeq
The last equation explains why the scalar fields $s$ and
$\chi$ are counted as $O(p^2)$ in the chiral expansion. Since $B_0
\neq 0$ in the chiral limit, the squares of the meson masses are
proportional to the quark masses contained in $s$ or $\chi$.
Having determined the two free constants $F,B_0$, one can now use
the effective Lagrangian (\ref{L2M}) or more precisely the generating
functional $Z_2[v,a,s,p]$ to derive all the current algebra results
for mesonic Green functions and amplitudes in terms of $F$ and
meson masses. The constant $B_0$ is hidden in the meson masses and
will never appear explicitly.

The calculation of Green functions and amplitudes to next--to--leading
order $p^4$ requires three steps \cite{GL1,GL2}~:
\bit
\item The general effective chiral Lagrangian ${\cL}_4$
to be considered at tree level.
\item The one--loop functional for the lowest--order Lagrangian
$\cL_2$ in (\ref{L2M}).
\item The Wess--Zumino--Witten functional \cite{WZW} to account for the
chiral anomaly \cite{ABJB}.
\eit
The general chiral Lagrangian $\cL_4(U,v,a,s,p)$ \cite{GL2},
\beqa
{\cL}_4
& = & L_1 \langle D_\mu U^\dagger D^\mu U\rangle^2 +
      L_2 \langle D_\mu U^\dagger D_\nu U\rangle
      \langle D^\mu U^\dagger D^\nu U\rangle \label{L4} \\*
& & + L_3 \langle D_\mu U^\dagger D^\mu U D_\nu U^\dagger D^\nu U\rangle +
    L_4 \langle D_\mu U^\dagger D^\mu U\rangle \langle \chi^\dagger U +
    \chi U^\dagger\rangle  \no \\
& & +L_5 \langle D_\mu U^\dagger D^\mu U(\chi^\dagger U + U^\dagger
    \chi)\rangle +
    L_6 \langle \chi^\dagger U + \chi U^\dagger \rangle^2 +
    L_7 \langle \chi^\dagger U - \chi U^\dagger \rangle^2  \no \\
& & + L_8 \langle \chi^\dagger U \chi^\dagger U +
 \chi U^\dagger \chi U^\dagger\rangle
    -i L_9 \langle F_R^{\mu\nu} D_\mu U D_\nu U^\dagger +
      F_L^{\mu\nu} D_\mu U^\dagger D_\nu U \rangle \no \\*
& & + L_{10} \langle U^\dagger F_R^{\mu\nu} U F_{L\mu\nu}\rangle +
    L_{11} \langle F_{R\mu\nu} F_R^{\mu\nu} + F_{L\mu\nu} F_L^{\mu\nu}\rangle +
    L_{12} \langle \chi^\dagger \chi \rangle~, \no
\eeqa
depends on 10 a priori arbitrary low--energy constants $L_1,\ldots,L_{10}$
($L_{11}$ and $L_{12}$ are not directly accessible experimentally).
These low--energy constants are the price we have to pay for using
an EFT instead of the fundamental Lagrangian (\ref{SM}).

Before discussing the low--energy constants further, let us consider
a general $L$--loop amplitude and determine its chiral dimension.
In this section, all internal lines correspond to pseudoscalar mesons
with propagators
\beq
\frac{i}{k^2 - M_a^2} \qquad \qquad a = 1,\ldots,8~.\label{Mprop}
\eeq
A general diagram contains $N_d^M$ vertices of $O(p^d)$ ($d = 2,4,\ldots)$
corresponding to the terms in the effective Lagrangian (\ref{LMeff}).
After the loop integrations, the amplitude is a homogeneous
function of the external momenta, the pseudoscalar masses $M_a$
and the external fields. The degree of homogeneity is called the
chiral dimension of the $L$--loop amplitude :
\beq
D_L = 4L - 2I + \sum_d dN_d^M,
\eeq
where $L$ is the number of loops and $I$ the number of internal lines.
Using the topological relation
\beq
L = I - \sum_d N_d^M + 1 ~,
\eeq
valid for any simply connected diagram, we obtain \cite{Wein79}
\beq
D_L = 2L + 2 + \sum_d (d-2) N_d^M, \qquad d = 2,4,\ldots \label{DLM}
\eeq

For a given $S$-matrix element, the chiral dimension $D_L$ increases with
$L$ according to Eq. (\ref{DLM}). In order to reproduce the (fixed)
physical dimension of the amplitude, each loop produces a factor $1/F^2$.
Together with the geometric loop factor $(4\pi)^{-2}$, the loop expansion
suggests
\beq
4\pi F \simeq 1.2 \mbox{ GeV}
\eeq
as the natural scale for the chiral expansion \cite{GeMa}.
We will have to check whether the coefficients of
the local Lagrangian $\cL_4$ in (\ref{L4}) respect this
scale.

It should not come as a surprise that a generic one--loop diagram
in CHPT is divergent (non--decoupling EFT are intrinsically
non--renormalizable). The divergences
can be taken care of once and for all by calculating the divergent
part of the one--loop functional. Since
it must be a local action with all the symmetries of $\cL_2$
\cite{Weinsy} and since it has chiral dimension $d=4$, the corresponding
Lagrangian must be of the general form (\ref{L4}), albeit with divergent
coefficients \cite{GL2}. The renormalization procedure at the one--loop
level is then achieved by simply dividing the coefficients in (\ref{L4})
into a (scale--dependent) finite  and a divergent part~:
\beq
L_i = L_i^r(\mu) + \Gamma_i \Lambda~. \label{Lisc}
\eeq
With dimensional regularization, the divergent factor $\Lambda$ is given by
\beq
\Lambda = \frac{\mu^{d-4}}{(4\pi)^2} \left\{\frac{1}{d-4} - \frac{1}{2}
[\log 4\pi + 1 + \Gamma'(1)]\right\}
\eeq
depending on the arbitrary renormalization scale $\mu$. The
relative coefficients $\Gamma_i$ reproduced in Table \ref{TabLi}
are of course chosen in such a way as to cancel the divergences
of the one--loop functional. The sum of one--loop amplitudes
and of tree amplitudes based on (\ref{L4}) is not only finite by
construction, but also independent of the scale $\mu$. Again, the
scale dependences of the finite one--loop amplitudes and of the
renormalized low--energy constants $L^r_i(\mu)$ in the tree
contributions cancel. That is all there is to renormalizing a
non--renormalizable EFT at the one--loop level~!

\begin{table}
\begin{center}
\caption{Phenomenological values and source for the renormalized coupling
constants $L^r_i(M_\rho)$ taken from Ref. \protect\cite{BEGRep}.}
\label{TabLi}
\vspace{.5cm}
\begin{tabular}{|c||r|l|r|}  \hline
i & $L^r_i(M_\rho) \times 10^3$ & source & $\Gamma_i$ \\ \hline
  1  & 0.7 $\pm$ 0.5 & $K_{e4},\pi\pi\rightarrow\pi\pi$ & 3/32  \\
  2  & 1.2 $\pm$ 0.4 &  $K_{e4},\pi\pi\rightarrow\pi\pi$&  3/16  \\
  3  & $-$3.6 $\pm$ 1.3 &$K_{e4},\pi\pi\rightarrow\pi\pi$&  0     \\
  4  & $-$0.3 $\pm$ 0.5 & Zweig rule &  1/8  \\
  5  & 1.4 $\pm$ 0.5  & $F_K:F_\pi$ & 3/8  \\
  6  & $-$0.2 $\pm$ 0.3 & Zweig rule &  11/144  \\
  7  & $-$0.4 $\pm$ 0.2 &Gell-Mann-Okubo,$L_5,L_8$ & 0             \\
  8  & 0.9 $\pm$ 0.3 & \small{$M_{K^0}-M_{K^+},L_5,$}&
5/48 \\
     &               &   \small{ $(2m_s-m_u-m_d):(m_d-m_u)$}       & \\
 9  & 6.9 $\pm$ 0.7 & $<r^2>_{em}^\pi$ & 1/4  \\
 10  & $-$5.5 $\pm$ 0.7& $\pi \rightarrow e \nu\gamma$  &  $-$ 1/4  \\
\hline
11   &               &                                & $-$1/8 \\
12   &               &                                & 5/24 \\
\hline
\end{tabular}
\end{center}
\end{table}

The final ingredient of $O(p^4)$ is the Wess--Zumino--Witten functional
to account for the chiral anomaly. Since I will not need it in the
following, let me only emphasize that it has no free parameters,
in contrast to the 10 parameters $L_i$ characterizing the ``normal''
Lagrangian (\ref{L4}).

\subsection{Low--Energy Constants}
The phenomenological determination of the constants $L_i$ was pioneered
by Gasser and Leutwyler \cite{GL2}. The present state is summarized in Table
\ref{TabLi}. If we compare the Lagrangians ${\cL}_2$ and
${\cL}_4$ and recall the chiral expansion parameter
\beq
\frac{p^2}{(4\pi F)^2}
\eeq
suggested by the loop expansion, we expect
\beq
L_i \; \lets \; \frac{1}{4(4\pi)^2} = 1.6 \cdot 10^{-3}~.
\eeq
Comparing with the phenomenological values $L_i^r(M_\rho)$ in
Table \ref{TabLi}, we conclude that this so--called naive chiral
power counting is  well satisfied in general, with the possible exception
of $L_3$, $L_9$ and $L_{10}$.

In principle, the low--energy constants $F$, $B_0$, $L_i$ should
be calculable
in QCD. Awaiting further theoretical progress, we may ask
the question whether the $L_i^r(M_\rho)$
can be understood in a more phenomenological way. In the spirit
of EFT, we expect all hadronic states that can couple to the pseudoscalar
mesons, but which are not included explicitly in the effective Lagrangian,
to contribute to the low--energy constants. Looking at the
hadronic spectrum, we therefore expect the meson resonances to
play an important r\^ole for understanding the numerical values
of the $L_i$.

A systematic analysis of the couplings between meson resonances of the
type $V$, $A$, $S$, $P$ and the pseudoscalar mesons was performed
in Ref. \cite{EGPR} (related work can be found in \cite{Resref}).
We use the general procedure discussed before of coupling matter fields
to the Goldstone bosons. For instance, an octet
\beq
R = \frac{1}{\sqrt{2}} \, \lambda_a R^a
\eeq
of resonance fields transforms as
\beq
R \stackrel{g}{\ra} h(g,\phi) R h(g,\phi)^{-1}
\eeq
where $h(g,\phi)$ is the compensator field in the triplet
representation of $SU(3)$. It turns out \cite{EGPR} that for $V$ and $A$
resonances only the octets can contribute to the $L_i$ with the
relevant couplings to the pseudoscalar mesons given by
\beqa
\cL_2^V & = & {F_V\over 2 \sqrt{2}}
     \langle V_{\mu\nu} f_+^{\mu\nu} \rangle +
    {iG_V\over \sqrt{2}} \langle V_{\mu\nu} u^\mu u^\nu \rangle \no \\*
\cL_2^A & = & {F_A \over 2 \sqrt{2}}
    \langle A_{\mu\nu} f_-^{\mu\nu} \rangle \label{L2res} \\*
f_\pm^{\mu\nu} & = & u F_L^{\mu\nu}u^\dg \pm u^\dg F_R^{\mu\nu}u~. \no
\eeqa
For $S$ and $P$, both octets and singlets can contribute.

The coupling constants $F_V$, $G_V$ and $F_A$ (and the corresponding ones for
$S$, $P$ resonances) can be estimated from resonance decays.
At $O(p^4)$, resonance exchange then contributes directly
to the $L_i$, e.g.
\beq
\ba{ll}
L_9^V = \dfrac{F_V G_V}{2M_V^2}~, \qquad & L_{10}^A = \dfrac{F_A^2}{4M_A^2}
\\[12pt]
M_V \simeq M_\rho~, & M_A \simeq M_{a_1}~.
\ea
\eeq
The results of Ref. \cite{EGPR} can be summarized as follows:
\begin{description}
\item[Chiral duality:] \mbox{ } \\
The $L_i^r(M_\rho)$ are practically saturated by resonance exchange.
There is very little room left for other contributions.
\item[Chiral VMD:] \mbox{ } \\
Whenever spin-1 resonances can contribute at all ($i = 1,2,3,9,10$),
the $L_i^r(M_\rho)$ are almost completely dominated by $V$- (and for
$L_{10}$ only, also $A$-) exchange. This also explains the relatively
large values of $L_3$, $L_9$ and $L_{10}$ in comparison with chiral
power counting.
\end{description}

As a practical consequence of chiral duality, all CHPT results to
$O(p^4)$ in the meson sector  can be recovered in the following way:
instead of the $O(p^4)$
Lagrangian (\ref{L4}) one can also use the lowest--order resonance
Lagrangian [the $V,A$ part is given in (\ref{L2res})] and employ
a cutoff $\mu\simeq M_\rho$ in the loop integrals. Of course, this
alternative approach bears the additional promise of making
sense also in higher orders of the chiral expansion (for a practical
application of this idea, see Ref. \cite{piBKM}).

Subsequently, it was shown \cite{EGLPR} that the
high--energy structure of QCD can provide additional information:
\bit
\item Imposing the QCD constraints at high energies via
dispersion relations,
all phenomenologically successful models for $V,A$ resonances were shown to
be equivalent to $O(p^4)$: tensor field description used in Eq.
(\ref{L2res}), massive Yang-Mills \cite{MeiPR}, hidden gauge formulations
\cite{Bando}, etc.
\item With additional QCD--inspired assumptions of
high--energy behaviour,
like an unsubtracted dispersion relation for the pion form factor, all
$V$ and $A$ couplings could be expressed in terms of $F_\pi$ and $M_\rho$
only. The results are in impressive agreement with the phenomenological
values of the $L_i^r(M_\rho)$ for $i = 1,2,3,9,10$.
\eit

There have been several attempts to calculate the $L_i$
``directly'' from QCD. The most complete and most advanced
analysis by Bijnens, Bruno and de Rafael \cite{BBR} is based on the
extended Nambu--Jona-Lasinio model. The earlier literature on
the subject can also be traced back from Ref. \cite{BBR}.


\subsection{Recent Applications}
Starting with the classical papers of Gasser and Leutwyler
\cite{GL1,GL2}, CHPT  has been applied to a large number of processes
to next--to--leading order. Referring to the existing
reviews [7--15] for a comprehensive survey of those results,
I will briefly mention some recent developments in the mesonic sector.

\noindent {\bf a. $K$ decays to $O(p^4)$} \\[5pt]
Inspired by the forthcoming facilities for high--intensity
kaon beams like the $\Phi$ factory DA$\Phi$NE \cite{DAF} in Frascati, both
semileptonic and non--leptonic $K$ decays have been studied
extensively. For semileptonic decays, the theoretical situation
is very favourable: all low--energy constants appearing in the decay
amplitudes are already known
(cf. Table \ref{TabLi}). Consequently, the CHPT amplitudes are fully
determined to $O(p^4)$ and will allow for non--trivial
tests of the standard model \cite{BEGRep}. The non--leptonic weak
interactions require a separate treatment. In particular, a
whole new set of low--energy constants appears \cite{KMW1,EKW} which
are far from being as well determined as the $L_i$ in the strong sector.
Fortunately, in rare $K$ decays re\-latively few of those constants
are relevant. In fact, some of the rare decays like
$K_S \to \gamma \gamma$ \cite{DAEG}, $K_L \to \pi^0 \gamma \gamma$
\cite{EPR2} or $K_S \to \pi^0 \pi^0 \gamma \gamma$ \cite{FK} are completely
independent of those constants. Although not as straightforward
as in the strong sector, the chiral anomaly also manifests
itself in non--leptonic kaon decays \cite{Kanom}.

\noindent {\bf b. Light quark mass ratios} \\[5pt]
The ratios of light quark masses can be determined to $O(p^4)$ \cite{GLPR}
by appealing to Dashen's theorem \cite{Dash} on the electromagnetic mass
shifts of pseudoscalar mesons. Recent investigations \cite{DHW}
have found substantial higher--order corrections
to Dashen's theorem leading in particular to a decrease of the ratio
$m_u/m_d$. A welcome consequence is that
the CHPT predictions for $\eta \to 3 \pi$ decay amplitudes are
increased as seems to be required by experiment.

\noindent {\bf c. CHPT beyond $O(p^4)$} \\[5pt]
In general, CHPT to $O(p^4)$ has fared quite well in comparison with
experiment. However, there are processes where higher--order
corrections seem to be larger than suggested by a naive application
of chiral dimensional analysis. One such process is $\gamma \gamma
\to \pi^0 \pi^0$ where the data \cite{CBC} show an enhancement
of the cross section near threshold. This enhancement can be
understood via a dispersion theoretical analysis of the scattering
amplitude \cite{Penn}. Very recently, the complete two--loop
calculation and estimates of the relevant low--energy constants
of $O(p^6)$ have become available \cite{BGS}. The resulting
cross section is in good agreement with experiment.

Another interesting transition is the non--leptonic decay $K_L \to \pi^0
\gamma \gamma$ where the experi\-mental spectrum in the invariant
mass of the two photons \cite{NA31} is in very good agreement
with the chiral prediction of $O(p^4)$ \cite{EPR2} while the experimental
rate seems to be at least twice as large as the theoretical value.
The dominant higher--order effects have recently been estimated \cite
{Kpgg}. Both unitarity corrections and vector meson exchange seem
to be required to understand both the rate and the spectrum
simultaneously.

\setcounter{section}{3}
\renewcommand{\theequation}{\arabic{section}.\arabic{equation}}
\setcounter{equation}{0}
\section{BARYONS}
The construction of chiral invariant Lagrangians with baryon fields
is in principle straightforward. As discussed in Sect. 2, there is a
well--defined procedure \cite{CCWZ} how to couple non--Goldstone
degrees of freedom in a chiral invariant manner. All we have to specify
is the transformation property of the respective fields under the
conserved subgroup.

\subsection{Relativistic Formulation}
Restricting ourselves to the case of two light flavours, we introduce the
familiar isodoublet nucleon field
\beq
\Psi = \left( \ba{cc} p \\ n \ea \right).
\eeq
Under chiral transformations, it transforms as
\beq
\Psi \stackrel{g}{\ra} \Psi' = h(g,\phi)\Psi \label{nlr}
\eeq
as discussed in Sect. 2.
The covariant derivative
\beqa
\nabla_\mu \Psi &=& \partial_\mu \Psi + \Gamma_\mu \Psi \\
\Gamma_\mu &=& \frac{1}{2} \{ u^\dg [\partial_\mu - i(v_\mu + a_\mu)]u
+ u [\partial_\mu - i(v_\mu - a_\mu)]u^\dg\} \no
\eeqa
now includes the external gauge fields $v_\mu$, $a_\mu$.

We now have all the ingredients at our disposal to construct the chiral
invariant Lagrangian for the pion--nucleon system
\beq
\cL_{\rm eff} = \cL_M + \cL_{MB}~. \label{Leff}
\eeq
As for the purely mesonic Lagrangian $\cL_M$ in (\ref{LMeff}),
we want to organize the meson--baryon Lagrangian $\cL_{MB}$ in a
systematic chiral expansion. Here we encounter a basic difference between
Goldstone and non--Goldstone fields. Since the nucleon mass remains
finite in the chiral limit, the four--momentum of a nucleon can never be
``soft''. This complicates the chiral counting considerably. For instance,
both $\Psi$ and $\nabla_\mu \Psi$ count as fields of $O(1)$, whereas
$(i \not\!\nabla - \stackrel{\circ}{m})\Psi$ is $O(p)$  like $u_\mu$
($\stackrel{\circ}{m}$ is the nucleon mass in the chiral limit).

We shall only consider the part of $\cL_{MB}$ that is bilinear in the
nucleon field. Because of the different Lorentz structure of meson and
baryon fields, the chiral expansion of $\cL_{MB}$ contains terms of
$O(p^n)$ for each positive integer $n$, unlike in the case of $\cL_M$
where the chiral expansion proceeds in steps of two powers of $p$.
In the two--flavour case we have \cite{GSS}
\beqa
\cL_{MB} &=& \cL_{\pi N}^{(1)} + \cL_{\pi N}^{(2)} + \cL_{\pi N}^{(3)}
+ \ldots   \label{LMB} \\
\cL_{\pi N}^{(1)} &=& \bar \Psi (i \not\!\nabla - \stackrel{\circ}{m} +
\frac{\stackrel{\circ}{g}_A}{2} \not\!u \gamma_5)\Psi~. \label{pN1}
\eeqa
The lowest--order Lagrangian $\cL_{\pi N}^{(1)}$ has only two parameters:
the nucleon mass $\stackrel{\circ}{m}$ and the neutron decay constant
$\stackrel{\circ}{g}_A$ (the superscript $\circ$ refers to the chiral $SU(2)$
limit). There are many more parameters in the higher--order Lagrangians
$\cL_{\pi N}^{(2,3)}$ \cite{GSS,Krause}. With the usual definition of the
$\pi N$ coupling constant $g_{\pi N}$ \cite{GSS},
the Goldberger--Treiman relation \cite{GoTr}
\beq
\stackrel{\circ}{g}_{\pi N} = \frac{\stackrel{\circ}{m}
\stackrel{\circ}{g}_A}{F}
\eeq
is an exact relation in the chiral limit. The formalism we have
used to construct chiral invariant Lagrangians automatically produces a
pseudo--vector $\pi N$ coupling. The derivative coupling makes a
fundamental property of Goldstone bosons manifest: their interactions
vanish as the energy goes to zero.

Starting with the work of Gasser, Sainio and \v Svarc \cite{GSS}, the
effective Lagrangian (\ref{Leff}) has been used to
calculate various Green functions and amplitudes with one incoming and
one outgoing baryon to one--loop accuracy (see Refs. \cite{RGass,RMei,MeiJMP}
for recent reviews). There is an intrinsic problem with chiral power
counting due to the presence of massive baryon lines in loop diagrams.
Referring to Ref. \cite{GSS} for a detailed analysis,
let me try to explain in a qualitative way the difference
between purely mesonic CHPT and CHPT with baryons. After the loop
integrations and after pulling out a common power of the meson
decay constant $F$,
a general CHPT amplitude is a homogeneous function of all the other
dimensional variables. Adopting a regularization procedure
that does not introduce an extrinsic scale like a momentum cutoff (in a
mass--independent renormalization scheme like dimensional
regularization with minimal subtraction, the dimensional scale appears only in
logarithms), the dimensional variables of a mesonic loop amplitude are
precisely the quantities that determine the chiral dimension: external
momenta, meson masses and external fields. This
leads to formula (\ref{DLM}) expressing the chiral dimension of the
amplitude in terms of the number of loops and of the chiral dimension
of vertices. In contrast, baryon propagators introduce the nucleon mass
as an additional dimensional quantity, which remains finite in the chiral
limit. Consequently, the degree of homogeneity of the baryonic amplitude
in question is no more related to its chiral dimension. Therefore, an
amplitude with given chiral dimension may receive contributions from
diagrams with an arbitrary number of loops \cite{GSS}.
In particular, the coupling
constants in the meson--baryon Lagrangian (\ref{LMB}) get
renormalized in every order of the loop expansion. Let us compare
once again with meson CHPT: the coupling constant $F$ in the
$O(p^2)$ mesonic Lagrangian (\ref{L2M}) is not renormalized\footnote{Remember
the constraint on the regularization procedure: a momentum cutoff would
lead to a spurious quadratic divergence in $F$ at one loop.}
because the divergence of any loop diagram corresponds to a local
Lagrangian with chiral dimension $d \geq 4$.

The nucleon mass is comparable to the intrinsic scale
$4\pi F \simeq 1.2$~GeV of CHPT. This suggests to set up baryon CHPT
in such a way as to expand in
$$
\frac{p}{4\pi F} \qquad \mbox{and} \qquad \frac{p}{\stackrel{\circ}{m}}
$$
simultaneously. In the relativistic formulation of CHPT \cite{GSS},
there is an essential difference between $F$ and $\stackrel{\circ}{m}$
in a generic
loop amplitude: $F$ appears only in the vertices, whereas the nucleon
mass is contained in the propagator.

\subsection{Heavy Baryon CHPT}
With inspiration from heavy quark effective theory \cite{HQET}, Jenkins
and Manohar \cite{JM1} have reformulated baryon CHPT in precisely such
a way as to transfer the nucleon mass from the propagators to the
vertices. Although sometimes
called the non--relativistic formulation of baryon CHPT, heavy baryon
CHPT is formulated in a Lorentz covariant way by defining
velocity--dependent fields \footnote{Following standard nomenclature,
both the external vector matrix field and the four-velocity
are denoted by the same symbol $v^\mu$.}
\beqa
N_v(x) &=& \exp[i\stackrel{\circ}{m}v \cdot x] P_v^+ \Psi(x) \label{vdf} \\
H_v(x) &=& \exp[i\stackrel{\circ}{m}v \cdot x] P_v^- \Psi(x) \no \\
P_v^\pm &=& \frac{1}{2} (1 \pm \not\!v)~, \qquad v^2 = 1~. \no
\eeqa
In the nucleon rest frame $v = (1,0,0,0)$ and $N_v$, $H_v$ correspond to the
usual non--relativistic projections of a Dirac spinor into upper--
and lower--component Pauli spinors. In addition, the exponential in
Eq. (\ref{vdf}) is designed to shift the dependence on the
nucleon mass. It is instructive to express the free Dirac Lagrangian
\beq
\cL_0 = \bar\Psi (i \not\!\partial - \stackrel{\circ}{m})\Psi
\eeq
in terms of the velocity--dependent fields $N_v$, $H_v$ as
\beq
\cL_0 = \bar N_v i v^\mu \partial_\mu N_v - \bar H_v (i v^\mu \partial_\mu
+ 2\stackrel{\circ}{m}) H_v + \bar N_v i\not\!\partial^\perp H_v +
\bar H_v i\not\!\partial^\perp N_v ~,
\eeq
with the transverse Dirac operator $\not\!\partial^\perp$ defined as
\beq
\not\!\partial = \not\!v v^\mu \partial_\mu  + \not\!\partial^\perp~.
\eeq
Ignoring the ``heavy'' field $H_v$ for the moment and thereby omitting
higher--order corrections in $1/\stackrel{\circ}{m}$, the large--components
field $N_v$ obeys the equation of motion
\beq
v^\mu \partial_\mu N_v = 0~,
\eeq
where the nucleon mass has disappeared by construction.
Therefore, the propagator of the field $N_v$
\beq
S(v \cdot k) = \frac{i P_v^+}{v \cdot k + i\ve} \label{Nprop}
\eeq
is independent of the nucleon mass, which must have migrated
somehow to the vertices of some effective Lagrangian. The momentum $k$ is
an off--shell momentum which is on the same footing as the off--shell
momentum of a pseudoscalar meson in the propagator (\ref{Mprop}).

Postponing the question what to do with the ``heavy''
field $H_v$, let us take a look at chiral power counting for a generic
loop diagram in heavy nucleon CHPT. The original meson--baryon
Lagrangian (\ref{Leff}) will be replaced by another effective Lagrangian
\beq
\cL'_{\rm eff} = \cL_M + \cL'_{MB}~, \label{Lp}
\eeq
where the meson Lagrangian $\cL_M$ is unchanged and the meson--baryon
Lagrangian $\cL'_{MB}$ to be constructed below depends only on $N_v$,
but not on $H_v$. Of course, it also depends on the meson fields via
$u(\phi)$ and on the various external fields. It is again organized in
terms with increasing chiral dimensions ($d = 1,2,3,\ldots$).

Because of the nucleon propagator (\ref{Nprop}) with chiral dimension
$-1$, the chiral dimension $D_L$ of a general $L$--loop amplitude with
nucleons and pions (more generally, baryons and pseudoscalar mesons)
is given by
\beq
D_L = 4L - 2 I_M - I_B + \sum_d d(N_d^M + N_d^{MB})~.
\eeq
The number of internal meson (baryon) lines is denoted by $I_M$ ($I_B$)
and the numbers of vertices of chiral dimension $d$ are called $N_d^M$,
$N_d^{MB}$ for the two parts of the effective Lagrangian (\ref{Lp}),
respectively. Specializing to the case of a single baryon line running
through the diagram\footnote{As we shall see shortly, there are no closed
baryon loops in this formalism to any finite order in
$1/\stackrel{\circ}{m}$.}, we have the relation
\beq
\sum_d N_d^{MB} = I_B + 1
\eeq
in addition to the general topological relation
\beq
L = I_M + I_B - \sum_d (N_d^M + N_d^{MB}) + 1~,
\eeq
valid for any simply connected diagram. Putting everything together,
we arrive at the analogue of relation (\ref{DLM}) in meson--baryon CHPT
with one external baryon:
\beq
D_L = 2L + 1 + \sum_d (d-2) N_d^M + \sum_d (d-1) N_d^{MB} \geq 2L+1~.
\label{DLB}
\eeq
In contrast to the relativistic formulation, chiral power
counting in the heavy baryon formulation is completely analogous to the
mesonic case. The chiral dimension increases with the number of loops.
As for the lowest--order constants $F$, $B_0$ in (\ref{L2M}), the
coupling constants in $\cL_{MB}'^{(1)} + \cL_{MB}'^{(2)}$ are not
renormalized in any order of the loop expansion, since $D_{L>0} \geq 3$.
This applies in particular to the lowest--order constants
$\stackrel{\circ}{m}$ and $\stackrel{\circ}{g}_A$.

The most transparent way to construct the effective meson--baryon
Lagrangian $\cL'_{MB}$ is to start from the path--integral representation
of the generating functional of Green functions in the relativistic
formulation \cite{GSS,MRR,BKKM}:
\beq
e^{iZ[j,\eta,\bar\eta]} = N \int [dU d\Psi d\bar\Psi]
\exp\left[ i \left\{ S_M + S_{MB} + \int d^4x(\bar\eta \Psi +
\bar\Psi \eta)\right\}\right]. \label{ZMB}
\eeq
The action $S_M + S_{MB}$ corresponds to the effective Lagrangian
(\ref{Leff}), the external fields $v$, $a$, $s$, $p$ are denoted
collectively as $j$ and $\eta,\bar\eta$ are fermionic sources. For two
light flavours, the pion--nucleon action is given by
\beq
S_{MB} = \int d^4x \left\{ \bar\Psi \left(i \not\!\nabla - \stackrel{\circ}{m}
+ \frac{\stackrel{\circ}{g}_A}{2} \not\!u \gamma_5\right)\Psi +
\cL_{\pi N}^{(2)} + \cL_{\pi N}^{(3)} + \ldots \right\}
\eeq
in accordance with Eqs. (\ref{LMB}),(\ref{pN1}). In terms of the
fields $N_v$, $H_v$ defined in (\ref{vdf}), the action takes the form
\beqa
S_{MB} &=& \int d^4x \{\bar N_v A N_v + \bar H_v B N_v +
\bar N_v \gamma^0 B^\dg \gamma^0 H_v - \bar H_v C H_v\} \\
A &=& P_v^+ \{ i v \cdot \nabla + \stackrel{\circ}{g}_A u \cdot S\}P_v^+
+ A^{(2)} + A^{(3)} + \ldots \\
B &=& P_v^- \left\{i \not\!\nabla^\perp - \frac{\stackrel{\circ}{g}_A}{2}
u \cdot v \gamma_5 \right\} P_v^+ + B^{(2)} + B^{(3)} + \ldots \\
C &=& P_v^- \{ iv \cdot \nabla + 2\stackrel{\circ}{m} + \stackrel{\circ}{g}_A
u \cdot S \} P_v^- + C^{(2)} + C^{(3)} + \ldots \\
\not\!\nabla^\perp &=& \not\!\nabla - \not\!v v \cdot \nabla ~. \no
\eeqa
The terms in $A$, $B$, $C$ with chiral dimension $d\geq 2$ are of course
due to the Lagrangian $\cL_{\pi N}^{(2)} + \cL_{\pi N}^{(3)} + \ldots$.
In $A$ and $C$, the spin operator
\beq
S^\mu = \frac{i}{2} \gamma_5 \sigma^{\mu\nu} v_\nu
\eeq
appears which obeys the relations \cite{JM1}
$$
S \cdot v = 0~, \qquad S^2 = - \frac{3}{4}~, \qquad
\{ S^\mu,S^\nu\} = \frac{1}{2} (v^\mu v^\nu - g^{\mu\nu})~,
$$
$$
[S^\mu,S^\nu] = i \ve^{\mu\nu\rho\sigma} v_\rho S_\sigma~.
$$
Rewriting also the source term in (\ref{ZMB}) in terms of the fields
$N_v$, $H_v$ with corresponding sources
\beq
\rho_v = P_v^+ \exp[i \stackrel{\circ}{m}v \cdot x] \eta~, \qquad
R_v = P_v^- \exp[i \stackrel{\circ}{m}v \cdot x] \eta~,
\eeq
we now integrate out the ``heavy'' components $H_v$. Shifting
variables
\beq
H'_v = H_v - C^{-1} B N_v
\eeq
and performing the Gaussian integration over the field $H'_v$, we arrive
at (dropping the sources $R_v$, $\bar R_v$ for simplicity)
\beq
e^{iZ[j,\rho,\bar\rho]} = N \int [dU dN_v d\bar N_v] \Delta_H
\exp\left[ i \left\{ S_M + S'_{MB} + \int d^4x (\bar \rho_v N_v +
\bar N_v \rho_v)\right\} \right]
\eeq
where
\beq
\Delta_H = \exp\left\{\frac{1}{2} \mbox{ tr } \log C \wh C^{-1}\right\} =
\exp\left\{ \frac{1}{2} \mbox{ tr } \log [1 + \wh C^{-1}(iv \cdot \Gamma
+ \stackrel{\circ}{g}_A u \cdot S + C^{(2)} + \ldots)]\right\}
\label{DelH} \eeq
is the corresponding determinant. Expanding the logarithm in (\ref{DelH})
and taking the trace, one obtains $\log \Delta_H$ as a sum of one--loop
functionals with an increasing number of vertices and free $H_v$
propagators $\wh C^{-1}$. On the other hand, the propagator
\beq
\wh C^{-1}(x) = (iv \cdot \partial + 2 \stackrel{\circ}{m} - i\ve)^{-1} =
i \exp[2i\stackrel{\circ}{m} v \cdot x] \Theta(-v \cdot x)
\int \frac{d^4k}{(2\pi)^3} \delta(k \cdot v)e^{-ikx} \label{Hprop}
\eeq
only describes backward propagation along the time--like
vector $v$ \cite{MRR}. This implies that the loops in the expansion of
$\Delta_H$ can never be closed in coordinate space.
Therefore, the trace in (\ref{DelH}) vanishes and
\beq
\Delta_H = 1~.
\eeq

Where have the closed baryon loops gone? Of course, they cannot have
disappeared because we have not made any approximations at all by
changing variables in the functional integral. The loops are in fact
still contained in the {\bf non--local} action
\beq
S'_{MB} = \int d^4x \bar N_v (A + \gamma^0 B^\dg \gamma^0 C^{-1} B)N_v~.
\eeq
In the corresponding determinant $\Delta_N$, both the propagators (\ref{Nprop})
(forward propagation along $v$) and (\ref{Hprop}) appear allowing for
non--vanishing traces in coordinate space.

At this point, the crucial approximation of heavy baryon CHPT is made:
the non--local part of the functional $S'_{MB}$ is expanded in a series
of local functionals with increasing chiral dimensions by expanding
$C^{-1}$ in a power series in $1/\stackrel{\circ}{m}$:
\beq
C^{-1} = P_v^- \left\{ \frac{1}{2\stackrel{\circ}{m}} -
\frac{iv \cdot \nabla + \stackrel{\circ}{g}_A u \cdot S}{(2\stackrel
{\circ}{m})^2} + O(p^2) \right\} P_v^- ~.
\eeq
At any finite order in $1/\stackrel{\circ}{m}$,
$S'_{MB}$ is a local action with terms of increasing chiral dimensions:
\beqa
\lefteqn{A + \gamma^0 B^\dg \gamma^0 C^{-1} B = A^{(1)} } \no \\
&& \mbox{} + A^{(2)} + \frac{1}{2\stackrel{\circ}{m}} \gamma^0 B^{(1)\dg}
\gamma^0 B^{(1)} \no \\
&& \mbox{} + A^{(3)} + \frac{1}{2\stackrel{\circ}{m}}
(\gamma^0 B^{(2)\dg} \gamma^0 B^{(1)} + \gamma^0 B^{(1)\dg} \gamma^0
B^{(2)}) - \frac{1}{(2\stackrel{\circ}{m})^2} \gamma^0 B^{(1)\dg}\gamma^0
(iv \cdot \nabla + \stackrel{\circ}{g}_A u \cdot S)B^{(1)} \no \\
&& \mbox{} + O(p^4) \label{Lexp}
\eeqa
with
\beqa
A^{(1)} &=& P_v^+ (iv \cdot \nabla + \stackrel{\circ}{g}_A u \cdot S)P_v^+
\label{AB} \\
B^{(1)} &=& P_v^- \left(i \not\!\nabla^\perp - \frac{\stackrel{\circ}{g}_A}{2}
u \cdot v \gamma_5\right) P_v^+ . \no
\eeqa
At the same time, closed fermion loops have disappeared from heavy baryon
CHPT. The only fermion propagator left in the theory is (\ref{Nprop})
which cannot give rise to closed loops.

We have now achieved our goal. The nucleon mass (or rather its chiral
limit value) has been transferred from the nucleon propagator to the
vertices of a local Lagrangian $\cL'_{MB}$ and it appears in exactly the
same way as the mesonic constant $F$. The scene is set for a systematic
chiral expansion of baryonic Green functions, simultaneously in
$p/4\pi F$ and $p/\stackrel{\circ}{m}$.

\subsection{Neutral Pion Photoproduction at Threshold}
Heavy baryon CHPT has been applied extensively (see
Refs. \cite{RMei,RMan} for reviews), not only to strong, electromagnetic
and semileptonic weak processes, but also to the non--leptonic  weak
interactions. Even in those cases where the loop calculation had
been done previously in the relativistic formulation, it is usually
instructive to redo the analysis in the framework of heavy
baryon CHPT \cite{BKKM}.

A nice example is neutral pion photoproduction at threshold
which was analyzed in the relativistic approach by Bernard, Gasser,
Kaiser and Mei\ss ner \cite{BGKM}. The process to be considered is
(see Ref.~\cite{David} for a general review of photo-- and
electroproduction of mesons)
\beq
\gamma(k) + p(p_1) \ra \pi^0(q) + p(p_2)
\eeq
and analogously for a neutron target. At threshold only the electric
dipole amplitude $E_{0+}$ contributes. It is related to the cross section
in the center--of--mass frame through
\beq
\lim_{\vec q \ra 0} \frac{|\vec k|}{|\vec q|} \frac{d\sigma}{d\Omega}
= (E_{0+})^2.
\eeq
To extract $E_{0+}$ from the amplitude calculated in heavy baryon CHPT,
it is convenient to write the scattering amplitude at threshold in the
lab frame (nucleon rest frame) as \cite{BKKM}
\beq
T_{\rm thresh} = 4\pi i \left( 1 + \frac{M_\pi}{m}\right) \chi_2^\dg
\vec \sigma \cdot \vec \ve(k) \chi_1 E_{0+} \label{thr}
\eeq
in terms of Pauli spinors $\chi_1$, $\chi_2$ and using the Coulomb
gauge $\ve^0(k) = 0$.

Let us now calculate the leading--order contribution to $T_{\rm thresh}$
in heavy baryon CHPT. Inspection of the effective Lagrangian
$\cL'_{MB}$ in Eq. (\ref{Lexp}) shows that the leading term (neither
$A^{(1)}$ nor $A^{(2)}$ contribute) is due to the $O(p^2)$ Lagrangian
\beq
\frac{1}{2 \stackrel{\circ}{m}} \bar N_v \gamma^0 B^{(1)\dg} \gamma^0
B^{(1)} N_v \label{lo2}
\eeq
with $B^{(1)}$ given in Eq. (\ref{AB}). The relevant terms in $B^{(1)}$
are obtained with
\beqa
\nabla^\mu &=& \partial^\mu + i e A^\mu \frac{1 + \tau_3}{2} + \ldots
\no \\
u^\mu &=& - \frac{1}{F} \partial^\mu \pi^0 \tau_3 + \ldots \no \\
v \cdot A &=& 0 \qquad \mbox{(Coulomb gauge)}
\eeqa
as
\beq
B^{(1)} = P_v^- (- e \not\!\!A \frac{1 + \tau_3}{2} +
\frac{\stackrel{\circ}{g}_A}{2F} v \cdot \partial \pi^0 \gamma_5 \tau_3)
P_v^+ . \label{B1}
\eeq
The scattering amplitude $T_{\rm thresh}$ can easily be read off from
Eqs. (\ref{lo2}) and (\ref{B1}):
\beqa
\frac{1}{2 \stackrel{\circ}{m}} \bar N_v \gamma^0 B^{(1)\dg} \gamma^0
B^{(1)} N_v  &
\wh{=}& \frac{-ie\stackrel{\circ}{g}_A}{4\stackrel{\circ}{m}F}
v \cdot q \bar N_v(\gamma^0 \not\!\ve^\dg \gamma^0 \gamma_5 +
\gamma^0 \gamma_5 \gamma^0 \not\!\ve) N_v = \no \\
&=& \frac{-i e \stackrel{\circ}{g}_A}{2 \stackrel{\circ}{m}F} v \cdot q
\bar N_v \not\!\ve \gamma_5 N_v =
\frac{-i e \stackrel{\circ}{g}_A}{\stackrel{\circ}{m}F} M_\pi \bar N_v
S \cdot \ve N_v~.
\eeqa
In the last step, use has been made of the threshold relation
$v \cdot q = M_\pi$ and of the identity
\beq
\bar N_v \gamma^\mu \gamma_5 N_v = 2 \bar N_v S^\mu N_v ~.
\eeq
In the lab frame
$$
S^0 = 0, \qquad \vec S = \frac{1}{2} \left( \ba{cc}
\vec \sigma & 0 \\ 0 & \vec \sigma \ea \right)
$$
and we recover the structure of Eq. (\ref{thr}) with
\beq
E_{0+}(\gamma p) = \frac{e \stackrel{\circ}{g}_A}{8\pi F}
\left[ \frac{M_\pi}{\stackrel{\circ}{m}} + O(p^2)\right] .
\eeq
The form of $B^{(1)}$ in (\ref{B1}) also shows that
$E_{0+}(\gamma n)$ vanishes at threshold in the same
leading order in CHPT. The threshold amplitudes are proportional to
the electric charges of the targets at leading order
in the chiral expansion.

At next--to--leading order, $O(p^3)$,
there are both tree--level and loop contributions to $T_{\rm thresh}$.
The obvious candidates for the tree--level amplitudes of $O(p^3)$ are
the terms in the third line of Eq. (\ref{Lexp}). However, it turns  out
that the corresponding amplitude vanishes at threshold \cite{BKKM}.
Incidentally, this already tells us that the loop amplitude must be
finite because chiral symmetry and gauge invariance do not permit
appropriate counterterms of $O(p^3)$.

On the other hand, there are non--vanishing Born diagrams with a nucleon
propagator between $O(p^2)$ $\gamma NN$ and $\pi NN$ couplings,
respectively. Those couplings originate both from the term proportional
to $\gamma^0 B^{(1)\dg} \gamma^0 B^{(1)}$ in (\ref{Lexp}) and from the
$O(p^2)$ Lagrangian \cite{BKKM}
\beq
A^{(2)} = \frac{1}{4\stackrel{\circ}{m}} P_v^+ \ve^{\mu\nu\rho\sigma}
v_\rho S_\sigma (c'_6 f_{+\mu\nu} + c'_7 \langle f_{+\mu\nu}\rangle )
P_v^+ + \ldots \label{AMM}
\eeq
with $f_{+\mu\nu}$ defined in Eq. (\ref{L2res}).
As is evident from the structure of the Lagrangian (\ref{AMM}), the
constants $c'_6$, $c'_7$ contribute to the Pauli form factor
$F_2^V(t)$ of the nucleons. In fact, these constants determine the
anomalous magnetic moments of the nucleons in the chiral limit
\cite{BKKM}:
\beq
\stackrel{\circ}{\kappa}_p \; = \frac{c'_6}{2} + c'_7~, \qquad \qquad
\stackrel{\circ}{\kappa}_n \; = - \frac{c'_6}{2} + c'_7 ~.
\eeq

Keeping track of the kinematic factor $1 + M_\pi/m$ in Eq. (\ref{thr}),
one arrives at the tree level amplitudes
\beqa
E^{\rm tree}_{0+}(\gamma p) &=& \frac{e g_A M_\pi}{8\pi F_\pi m}
\left[ 1 - \frac{M_\pi}{2m}(3 + \kappa_p) + O(p^2)\right] \label{Etree} \\
E^{\rm tree}_{0+}(\gamma n) &=& \frac{e g_A M_\pi}{8\pi F_\pi m}
\left[ \frac{M_\pi \kappa_n}{2m} + O(p^2)\right].  \no
\eeqa

The alert reader will have noticed that I have used in Eq. (\ref{Etree})
the physical values
$g_A$, $m$, $F_\pi$, $\kappa_p$, $\kappa_n$ instead of the chiral limit
values. That I could do so with impunity, is a
major conceptual advantage of heavy baryon CHPT. The relation
between the physical and the chiral limit values of all those quantities
is such that the differences can be shoved into the higher--order
contributions denoted by $O(p^2)$ in (\ref{Etree}). This is not at all the
case in the relativistic formulation for the reasons already discussed.
In fact, most of the loop contributions encountered in the relativistic
approach renormalize the various constants to their physical values
\cite{BGKM}. The final result is the same in both approaches,
but the amount of work needed is quite different.

This brings us to the final and most important point. As first shown in
Ref. \cite{BGKM}, there are in addition to the tree--level amplitudes
(\ref{Etree}) non--trivial loop contributions to the electric dipole
amplitudes at threshold at next--to--leading order in the chiral
expansion. Once again, it is much easier in the heavy baryon formulation
to locate the relevant diagrams \cite{BKKM}. According to chiral power
counting as expressed in Eq. (\ref{DLB}), the one--loop contributions
of $O(p^3)$ only involve vertices of the $O(p)$ Lagrangian
$A^{(1)}$ in (\ref{AB}). Consequently, there is no direct $\gamma NN$
coupling in the Coulomb gauge $v \cdot A = 0$. Moreover, at threshold
there is also no direct coupling of the produced $\pi^0$ to the nucleon
because the coupling strength $S \cdot q$ vanishes at threshold (in the
nucleon rest frame, $S^0 = 0$ and $\vec q = 0$ at threshold). With these
simplifications available in the heavy baryon formulation, out of some
60 diagrams only four non--trivial ones survive, the so--called triangle
and rescattering diagrams and their crossed partners \cite{BKKM}. As
noticed before, the loop amplitudes are finite and they contribute
equally to $E_{0 +}(\gamma p)$ and $E_{0+}(\gamma n)$.

The loop amplitudes were omitted in the original derivations \cite{VZB}
and many later re\-derivations of the low--energy theorems for the
electric dipole amplitudes at threshold. The complete low--energy
theorems take the form \cite{BGKM}
\beqa
E_{0+}(\gamma p) &=& \frac{e g_A M_\pi}{8\pi F_\pi m}
\left[ 1 - \frac{M_\pi}{2m}\left(3 + \kappa_p + \frac{m^2}{8 F^2_\pi}\right)
+ O(p^2)\right] \label{Etot} \\
E_{0+}(\gamma n) &=& \frac{e g_A M_\pi}{8\pi F_\pi m}
\left[ \frac{M_\pi}{2m} \left(\kappa_n - \frac{m^2}{8F^2_\pi}
\right) + O(p^2)\right].  \no
\eeqa
Although there still seems to be some confusion in the literature
\cite{SKF} as to the status of these results,
the low--energy theorems (\ref{Etot}) are exact predictions of
QCD in the limit of isospin conservation. It is a different story
that it may be difficult to verify the low--energy theorems
experimentally. The relativistic calculation suggests \cite{BGKM,BKM1}
that the chiral expansion for $E_{0+}$ may be slowly converging.
Moreover, it may be necessary to include isospin violation \cite{BKM2}
to extract the threshold amplitudes from experiment.

\vspace{1cm}
\noindent {\bf Acknowledgements}

\noindent I want to thank J\"urg Gasser and Helmut Neufeld for helpful
comments.

\newpage

\newcommand{\PL}[3]{{Phys. Lett.}        {#1} {(19#2)} {#3}}
\newcommand{\PRL}[3]{{Phys. Rev. Lett.} {#1} {(19#2)} {#3}}
\newcommand{\PR}[3]{{Phys. Rev.}        {#1} {(19#2)} {#3}}
\newcommand{\NP}[3]{{Nucl. Phys.}        {#1} {(19#2)} {#3}}

\end{document}